\documentclass[useAMS,usenatbib,usegraphicx]{mn2e}
\usepackage{amsmath}
\usepackage{color}

\newif\ifAMStwofonts
\AMStwofontstrue

\newcommand{\simlt}{\lower.5ex\hbox{$\; \buildrel < \over \sim \;$}}
\newcommand{\simgt}{\lower.5ex\hbox{$\; \buildrel > \over \sim \;$}}

\title[Assembly bias in close pairs]
{Assembly bias evidence in close galaxy pairs}
\author[Ferreras et al.]
{I. Ferreras$^{1,2,3,4}$\thanks{E-mail: iferreras@iac.es},
A.~M. Hopkins$^5$, C. Lagos$^6$, A.~E. Sansom$^7$, N. Scott$^{8,9}$, \and
S.~M. Croom$^{8,9}$, S. Brough$^{10}$
\\
$^1$ Mullard Space Science Laboratory, University College London, 
Holmbury St Mary, Dorking, Surrey RH5 6NT, UK\\
$^2$ Department of Physics and Astronomy, University College London,
Gower Street, London WC1E 6BT, UK\\
$^3$ Instituto de Astrof{\'i}sica de Canarias, Calle V{\'i}a L{\'a}ctea s/n,
E38205, La Laguna, Tenerife, Spain\\
$^4$ Departamento de Astrof{\'i}sica, Universidad de La Laguna (ULL), E-38206 La Laguna,
Tenerife, Spain\\
$^5$ Australian Astronomical Optics, Macquarie University, 105 Delhi Rd, North Ryde, NSW 2113, Australia\\
$^6$ ICRAR, M468, University of Western Australia, 35 Stirling Hwy, Crawley, WA 6009, Australia\\
$^7$ Jeremiah Horrocks Institute, University of Central Lancashire, Preston PR1 2HE, UK\\
$^8$ Sydney Institute for Astronomy, School of Physics, A28, The University of Sydney, Sydney, NSW 2006, Australia\\
$^9$ ARC Centre of Excellence for All Sky Astrophysics in 3~Dimensions (ASTRO 3D)\\
$^{10}$ School of Physics, University of New South Wales, NSW 2052, Australia\\
}

\begin{document}
\date{MNRAS, in press. Accepted 2019 May 6. Received 2019 March 20; in original form 2018 December 17}
\pagerange{\pageref{firstpage}--\pageref{lastpage}} \pubyear{2019}
\maketitle
\label{firstpage}


\begin{abstract}
The growth channel of massive galaxies involving mergers can be
studied via close pairs as putative merger progenitors, where the
stellar populations of the satellite galaxies will be eventually
incorporated into the massive primaries. We extend our recent analysis
of the GAMA-based sample of close pairs presented in Ferreras et
al. to the general spectroscopic dataset of SDSS galaxies (DR14), for which the high
S/N of the data enables a detailed analysis of the differences between
satellite galaxies with respect to the mass of the primary galaxy.
A sample of approximately two thousand satellites of massive galaxies
is carefully selected within a relatively narrow redshift range (0.07$<$z$<$0.14).
Two main parameters are considered as major drivers of the star formation history
of these galaxies, 
namely: the stellar velocity dispersion of the satellite ($\sigma$), as a
proxy of ``local'' drivers, and the ratio between the stellar mass of
the satellite and the primary, $\mu=M_{\rm SAT}/M_{\rm PRI}$, meant to
serve as an indicator of environment.  Consistently with the
independent, GAMA-based work, we find that satellites around the most
massive primaries appear older, at fixed velocity dispersion, than
satellites of lower mass primaries. This trend is more marked in
lower mass satellites ($\sigma$$\sim$100\,km\,s$^{-1}$),
with SSP-equivalent age differences up to
$\sim$0.5\,Gyr, and can be interpreted as a one-halo assembly bias, so that
satellites corresponding to smaller values of $\mu$ represent older
structures, akin to fossil groups.
\end{abstract} 


\section{Introduction}
\label{Sec:Intro}

A wide range of factors determine the formation of galaxies, as they
evolve from clumps of gas that follow the original density
fluctuations at early times to the complex web of galaxies that we see
today. One can separate potential mechanisms that influence galaxy
formation and evolution between local (“short-range" mechanisms that
extend over $\sim$1-10\,kpc scales) and global ones
(environment-related processes that can affect galaxy formation over
much larger scales). Alternatively, one can look into the formation of
galaxies as a ``two stage'' process \citep{Oser:12}, distinguishing
between an in-situ formation phase -- from the collapse of gas and
subsequent cooling into the newly forming galaxy -- and an ex-situ
component made up of stars formed in other galaxies, accreted through
merging. This split is especially informative in massive galaxies
(defined as those having a stellar mass higher than $\sim
10^{11}$M$_\odot$), as the massive cores of these galaxies
feature old, metal-rich and [$\alpha$/Fe] overabundant
populations \citep[e.g.][]{Thomas:05,IGDR:11}, as expected from an
early, strong, and short-lived episode of formation.
This pattern of
ages and abundances in massive galaxies is suggestive of 
a substantial contribution from the in-situ phase.
In contrast, numerical simulations indicate that the outer envelope
may be built up through the ex-situ growth
channel \citep[e.g.][]{Naab:09}.
Radial gradients of stellar
populations in massive galaxies, therefore, allow us to understand the
role of these two phases in the formation of massive galaxies at
present time. For instance, early-type galaxies show a strong negative
metallicity gradient but a relatively shallow age
profile \citep[e.g.][]{FLB:12,Greene:15,Goddard:16}.  The outer
regions of massive galaxies also have enhanced [alpha/Fe] ratios,
indicative of short star-formation timescales, ceasing at early
times \citep{Greene:13,Greene:15}. Therefore, taking the outer regions
of massive galaxies as constituted by the ex-situ phase, we would
infer that the progenitors that feed this phase are low-metallicity
(i.e. potentially low mass) galaxies formed at early times. Is it
possible to view this external envelope dein the making?

Observations of close pairs allow us to probe the progenitors of
eventual galaxy mergers \citep[e.g.][]{Patton:00,Lin:04,SolAlonso:04,Ellison:08,
Rogers:09,CLS:12,EMQ:12}.
By choosing a relatively small separation,
both spatially (in projection) and in velocity (along the line of sight),
it is possible to identify systems dynamically bound and potentially
merging. This technique can be exploited to determine merger
rates, assess feedback effects  and study the properties
of the progenitors that will form the future, merged, system
(see the above references for a non-extensive selection of results).
In \citet{SH4} a sample of z$\simlt$1.5 massive galaxies
was selected from the SHARDS deep, medium-band
survey \citep{SHARDS}  to assess the growth of the ex-situ phase as a function of the
merger ratio. A dominant contribution was found from mergers with satellite-to-primary mass ratio 
in the range
$\mu\equiv M_{\rm SAT}/M_{\rm PRI}=[0.5,1]$, i.e. putting the contribution of
minor mergers in a subdominant category. 
This result was found to hold at lower redshift, as presented
in the SDSS-based study of \citet{Ruiz:14}.
Note that we should not extrapolate this trend to the general population of
(lower mass) galaxies, where minor merging can dominate
the ex-situ phase \citep[e.g.][]{Sugata:14}. A follow-up study
selecting close pairs involving, at least, a massive
galaxy was undertaken in the Galaxy And Mass Assembly (GAMA)
survey \citep{GAMA}, where the high
completeness of the survey and the availability of optical
spectroscopy -- mostly from the Anglo-Australian Telescope (AAT) --
allowed us to perform an
unbiased search of variations in the stellar populations of the
merger progenitors \citep{GAMA:CPs}. An intriguing segregation was found so that,
at fixed stellar mass, satellites orbiting the most massive galaxies
were, on average, 1\,Gyr older than those around the lower mass
primaries. This effect was found not to depend on whether the
stellar mass of the primary or the hosting halo mass is used
to split the sample. This result is reminiscent of galactic
conformity, i.e. the tendency for the age of a galaxy to “align” with
the age of the corresponding central galaxy \citep{Weinmann:06,Hartley:15,Kawinwanichakij:16},
and leads to the
concept of galaxy conformity \citep[see][for a recent review]{WT:18}.
A common interpretation of galaxy conformity is that it could
be driven by assembly bias \citep{Gao:05}, that refers to the
effect of halos in higher density environment being older and more
concentrated at fixed stellar mass \citep{Hearin:15}. In the latter
scenario, galaxy properties would depend on halo properties beyond its
virial mass.  There are contradictory results in the literature as to
whether assembly bias is at the heart of galaxy conformity
\citep[e.g.,][]{Treyer:18} or is due to internal physical processes in
halos that are unrelated to halo age \cite[e.g.,][]{ZM:18}.  In
simulations, however, the connection between conformity and assembly
bias is more clearly established \citep[e.g.,][]{Bray:16,Paranjape:15,PP:17}. 
Observational studies of this bias provide valuable clues about the
interplay between structure formation, driven by dark matter, and
galaxy formation, which strongly depends on a plethora of physical
processes termed `baryon physics'.

\begin{figure}
\begin{center}
\includegraphics[width=88mm]{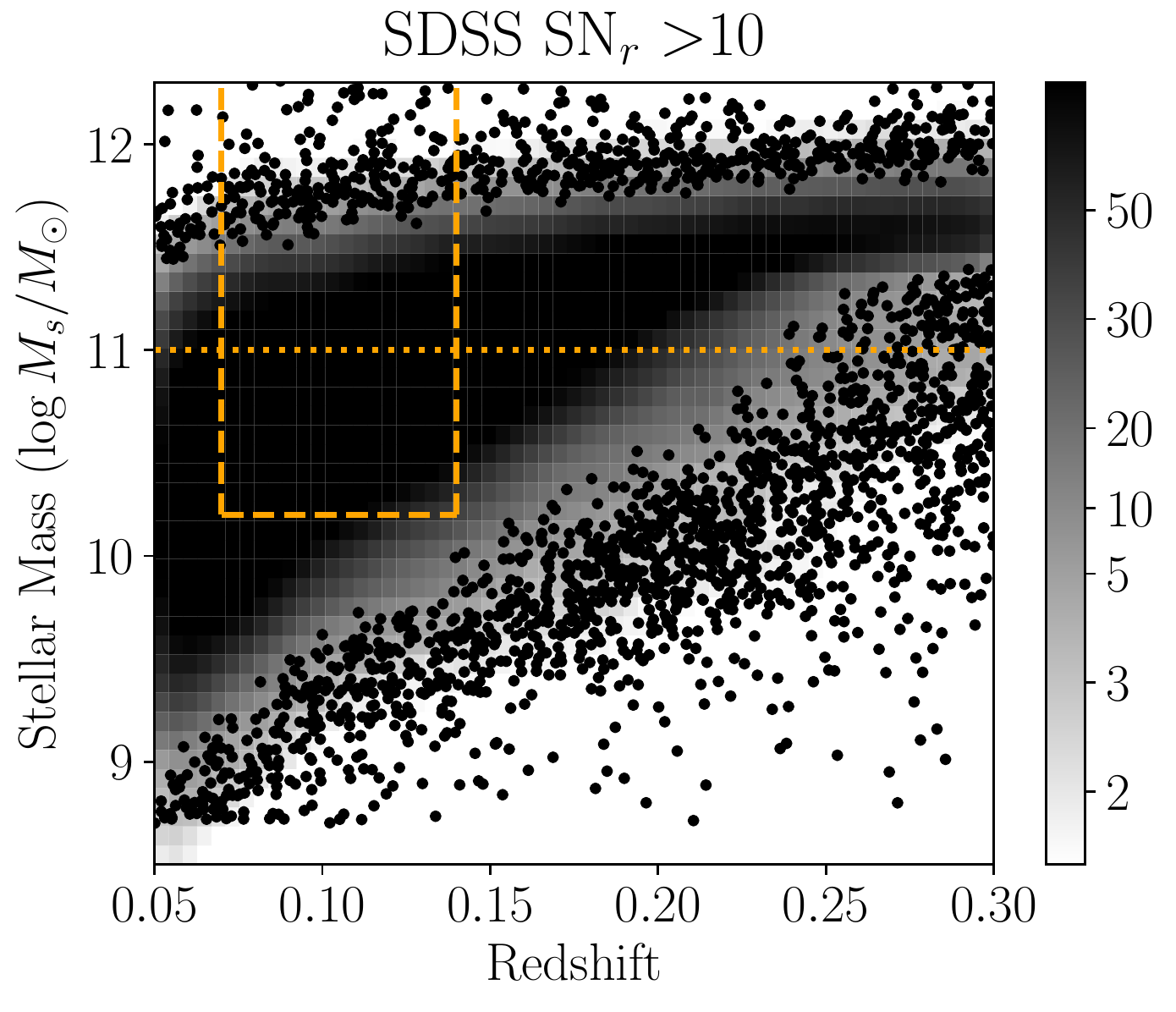}
\end{center}
\caption{Density plot showing the number of available SDSS
spectra on a stellar mass vs redshift diagram. The greyscale
is applied to the region of high density of spectra, whereas in the
lower density regions the individual data points are shown as
dots. The dashed lines mark the region from which the final 
sample of close pairs is extracted (see Fig.~\ref{fig:sample} and text for 
details). For reference, all primary galaxies are
located above the horizontal dotted line.}
\label{fig:MassComplete}
\end{figure}

In this paper we extend the analysis presented in \citet{GAMA:CPs} by
exploring an independent sample from the Sloan Digital Sky
Survey \citep[SDSS,][]{SDSS}. This work, therefore, focuses on close pairs involving
at least a massive galaxy, and the potential effect that a massive
primary, or its environment, could exert on the associated satellite. 
The larger data volume probed by the SDSS and the exquisite flux calibration
allow us to revisit the question of whether there are substantial differences
in the stellar populations of satellites orbiting massive galaxies and
to produce a more accurate estimate of the stellar age differences, and
a more comprehensive analysis of possible biases due to the sample selection.
A standard $\Lambda$CDM cosmology is
adopted, with $\Omega_m=0.27$ and
H$_0=70$\,km\,s$^{-1}$Mpc$^{-1}$. For reference, the look-back time to
z=0.1 (roughly the median of our working sample) is 1.30\,Gyr and
the 3\,arcsec diametre fibre of the SDSS spectrograph maps into
a projected distance of 5.5\,kpc at that redshift.

\section{SDSS sample selection}
\label{Sec:Data}

We retrieve from the Sloan Digital Sky Survey (SDSS)
DR14 archive \citep{SDSSDR14} all spectra from the
classic SDSS database, classified as a galaxy, in the redshift range
$0.05<z<0.3$, and with a SNR in the $r$ band above 10.0. Moreover, we
reject data with a raised zWARNING flag. The spectra was
cross-matched with the Johns Hopkins/MPA catalogue to retrieve
the stellar masses, based on methods set out in \citet{Kauff:03}.
The resulting sample comprises 531,280 spectra.
Fig.~\ref{fig:MassComplete} illustrates the
stellar mass completeness of the full set, by showing the
number of available SDSS galaxy spectra on a diagram of stellar
mass versus redshift (in regions with lower numbers of available
spectra the shaded representation is substituted by individual data
points). The dashed lines delimit the range in redshift and stellar
mass probed in this sample (see below).
From this sample, we select all massive galaxies, defined as having a stellar mass
above $10^{11}$M$_\odot$ (i.e. above the horizontal dotted line in Fig.~\ref{fig:MassComplete}).
The resulting sample of 186,824 massive
galaxies is then searched for the availability of SDSS spectra of nearby galaxies --
selected from the parent sample of high SNR data -- located
within a projected radius of 100\,kpc and with a peculiar velocity,
derived from the redshift difference, within $\pm 700$\,km/s. We note
this is the same criterion applied to our GAMA-based sample of close
pairs, as presented in \citet{GAMA:CPs}.

\begin{figure}
\begin{center}
\includegraphics[width=84mm]{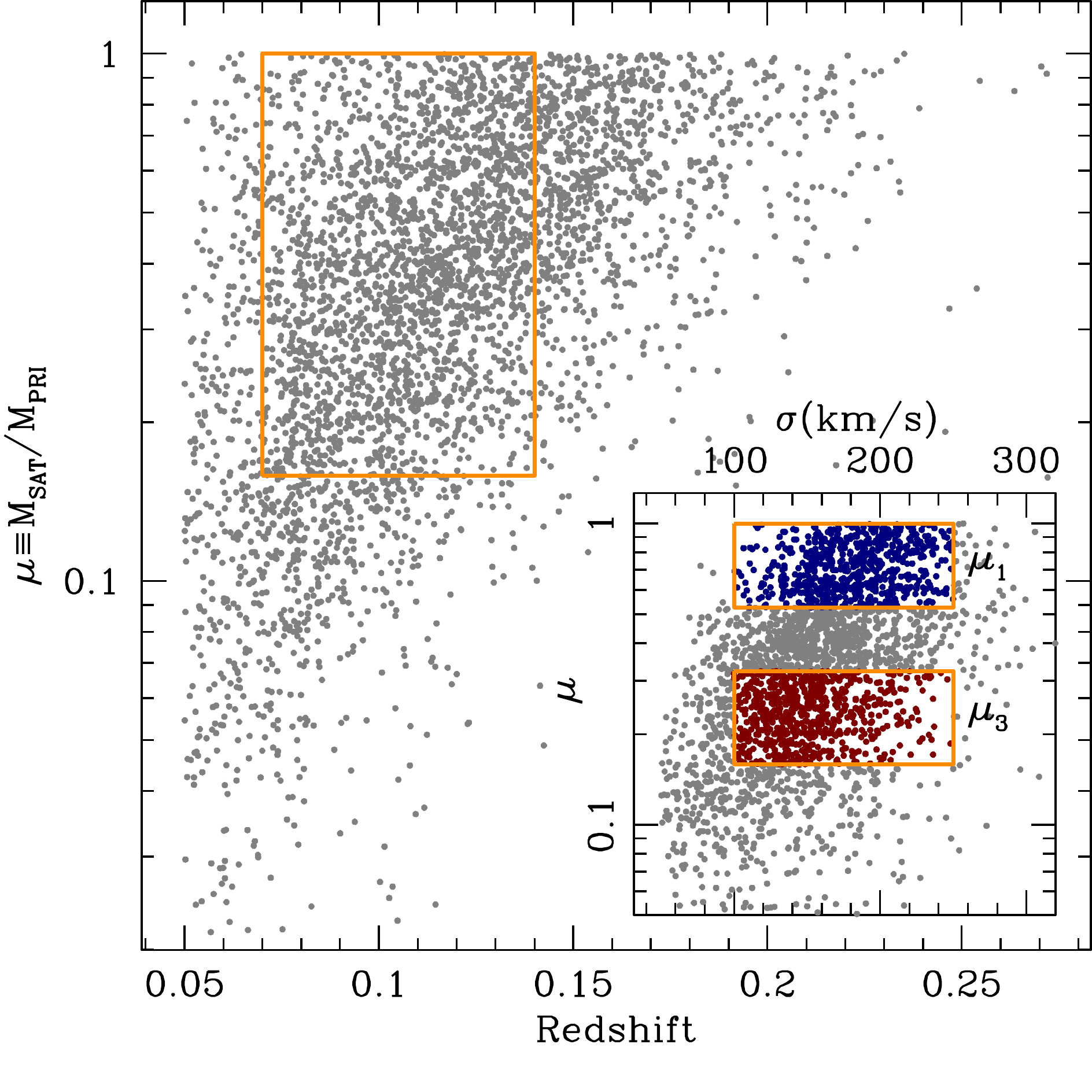}
\end{center}
\caption{This figure illustrates the stacking strategy.
The grey dots show the distribution of redshift vs stellar mass ratio
for the starting sample of close pair systems. The orange box encloses
the narrower redshift interval (0.07$<$z$<$0.14) over which we can
probe a wide range of mass ratios ($\log\mu\geq -0.8$). The inset
shows the sample within this redshift and mass ratio window, with
respect to velocity dispersion, our ``local proxy''.  The interval
100$<$$\sigma$$<$250\,km\,s$^{-1}$ is further divided into three
terciles with respect to the mass ratio, from which we extract the
highest (i.e. major merger progenitor) and the lowest (i.e. minor
merger progenitor) terciles, as shown in blue (top box, $\mu_1$) and
red (bottom box, $\mu_3$), respectively.}
\label{fig:sample}
\end{figure}

Fig.~\ref{fig:sample} shows the general sample on a diagram plotting
the stellar mass ratio ($\mu\equiv M_{\rm SAT}/M_{\rm PRI}$) vs
redshift. The selection effect is apparent, with a strong Malmquist
bias towards higher masses at higher redshift, as expected since the
SDSS spectroscopic survey is limited to $r<$17.7\,AB~mag. Moreover, our
high S/N constraint accentuates the effect of the bias.
Therefore, a
naive adoption of all spectra as shown by the grey dots will
introduce a systematic, such that at
high $\mu$, we would be selecting a wide range of
redshifts, whereas satellite spectra corresponding to
low $\mu$ only probe the lower redshift subset. Furthermore, the wide
redshift covered introduces an additional bias due to the fixed
aperture imposed by the 3\,arcsec fibres of the SDSS
spectrograph. Over the full z=0.05--0.3 redshift interval shown in the
figure, the fibre maps a
physical size between 2.9 and 13.3\,kpc. The presence of population
gradients will therefore produce an additional systematic trend.

In order to mitigate these possible biases, we restrict the redshift
range to z=0.07--0.14 (cyan dashed lines in Fig.~\ref{fig:MassComplete}, 
and orange box in Fig.~\ref{fig:sample}).
Note that within the range of redshift and stellar mass
probed here, we do not expect an incompleteness from the flux limit of the parent sample.
The inset in Fig.~\ref{fig:sample} 
shows our working sample on a diagram with stellar mass ratio vs
velocity dispersion. We impose a further constraint by restricting the
sample in velocity dispersion between 100 and 250 km\,s$^{-1}$,
splitting the interval into five equal steps of width
$\Delta\sigma$=30\,km\,s$^{-1}$.  We choose galaxies with mass ratios
$\log\mu>-0.8$, and split the sample into three terciles, where the
highest and lowest bins are colour coded in blue ($\mu_1$) and red ($\mu_3$),
respectively.  Table~\ref{tab:stacks} shows the details of the
stacks. Taking all galaxies within our working sample regardless of
the stellar mass ratio, i.e. the $\mu_0$ subset in
Table~\ref{tab:stacks}, we obtain a median redshift $z_M=0.11\pm
0.02$, which implies a variation in the physical extent of the SDSS
spectroscopic fibre of $\sim\pm$1\,kpc.

\begin{figure}
\begin{center}
\includegraphics[width=80mm]{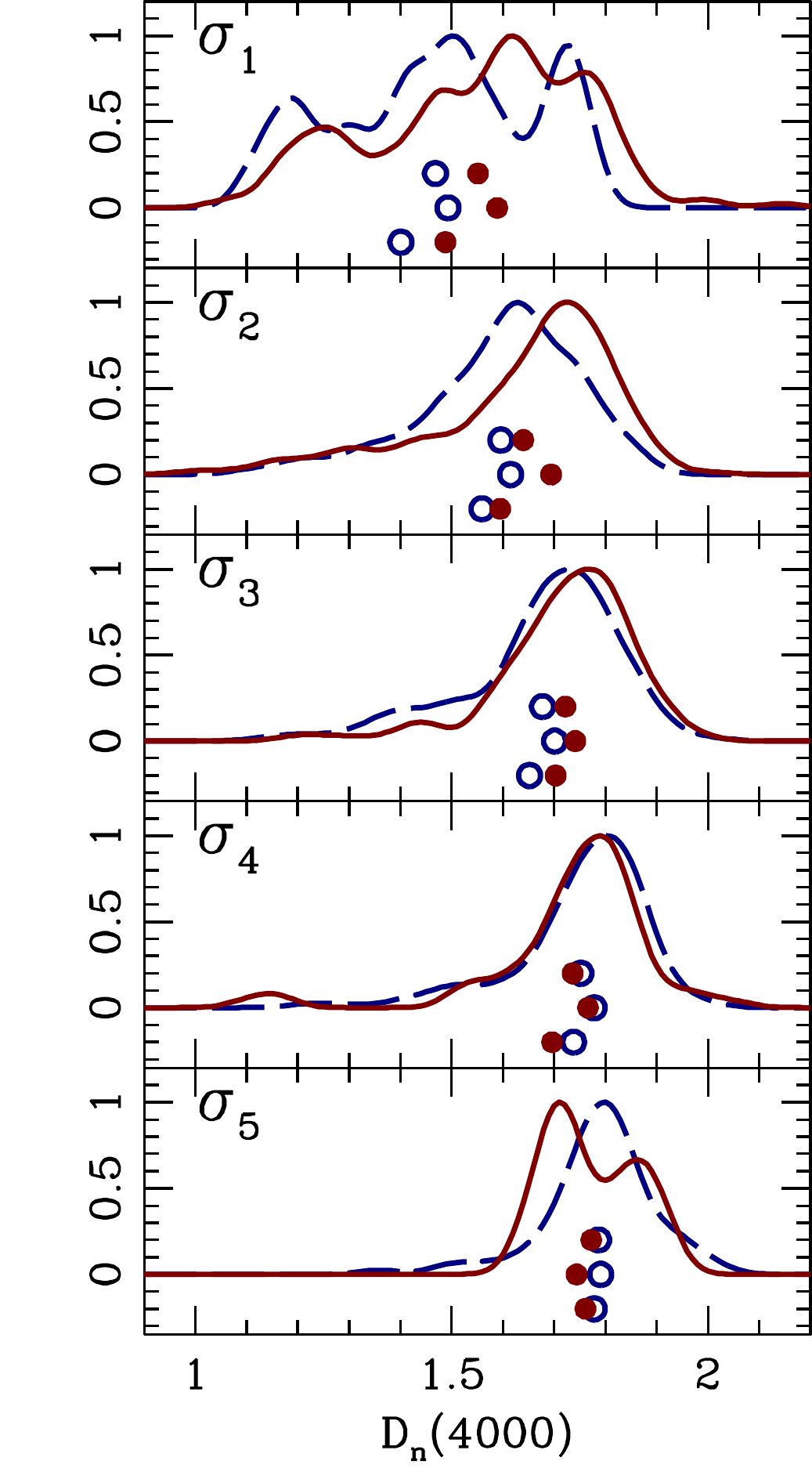}
\end{center}
\caption{Histograms with the 4000\AA\ break strength
measured in {\sl individual} spectra. The histograms are renormalized
within each bin ($\sigma_1\cdots\sigma_5$).  The solid red lines
correspond to satellites around the most massive primaries
(i.e. subset $\mu_3$), and the dashed blue lines represent satellites
around lower mass primaries (i.e. subset $\mu_1$). The dots, from bottom to top are the Dn(4000) measurements of
the stacked spectra; the median, and the mean of the individual
estimates, respectively. In each case, the blue open dots (red solid dots) represent
the $\mu_1$ ($\mu_3$) subsample.}
\label{fig:Dn4K}
\end{figure}

\begin{table*}
\caption{Number and S/N (in brackets) of spectra used in the stacks
(see Fig.~\ref{fig:sample}). The S/N is given per pixel, averaged in
the 5,000-5,500\AA\ spectral window. The $\sigma_1\cdots\sigma_5$ cases
represent the bins regarding the velocity dispersion of the satellite
galaxy (with the interval quoted underneath in km/s).  }
\label{tab:stacks}
\begin{tabular}{cr|rc|ccccc}
\hline
 & & & & \multicolumn{5}{c}{Number of spectra (S/N)}\\
\hline
 & & & & $\sigma_1$ & $\sigma_2$ & $\sigma_3$ & $\sigma_4$ & $\sigma_5$\\
ID & $\log\mu\equiv\log M_{\rm SAT}/M_{\rm PRI}$ & N & z$_M$ & 100-130 & 130-160 & 160-190 & 190-220 & 220-250\\
\hline
$\mu_1$ & $(-0.278, 0.000]$ &  663 & $0.11799\pm 0.01918$ &  39 ( 73) & 145 (154) & 215 (203) & 176 (205) &  88 (157)\\
$\mu_2$ & $(-0.490,-0.278]$ &  669 & $0.11143\pm 0.01846$ & 110 (114) & 231 (195) & 190 (192) &  82 (135) &  56 (114)\\
$\mu_3$ & $[-0.800,-0.490]$ &  672 & $0.10038\pm 0.01801$ & 189 (155) & 258 (199) & 130 (156) &  71 (123) &  24 ( 75)\\
\hline
$\mu_0$ & $[-0.800, 0.000]$ & 2004 & $0.10934\pm 0.01933$ & 338 (201) & 634 (316) & 535 (313) & 329 (269) & 168 (207)\\
\hline
\end{tabular}
\end{table*}

\begin{figure}
\begin{center}
\includegraphics[width=8.3cm]{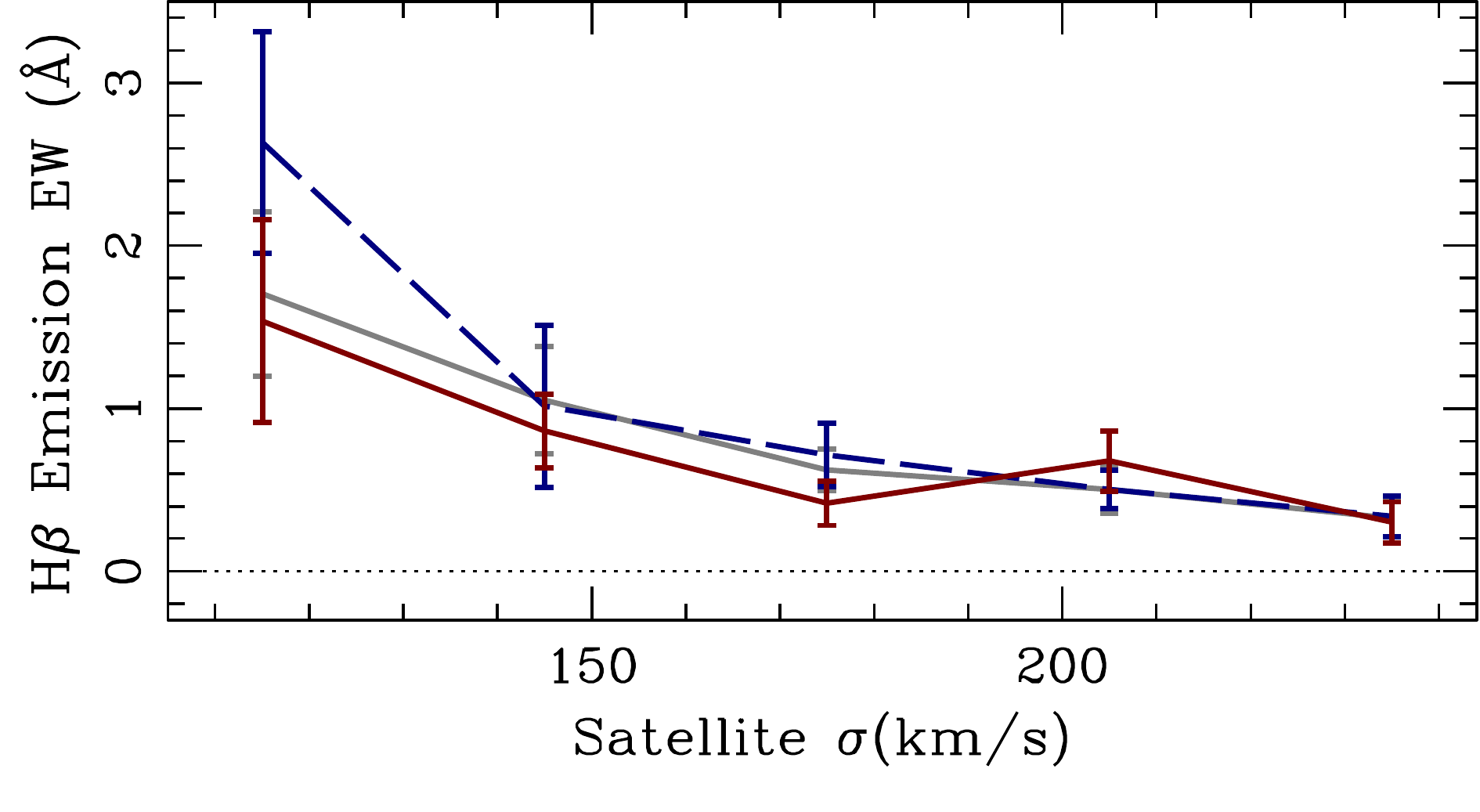}
\end{center}
\caption{Correction for emission in the H$\beta$ line,
measured as an equivalent width. The error bars have
been blown up by a factor of $3$. The solid red (dashed blue)
lines correspond to satellites around the most (least)
massive primaries. The grey line is the trend for the
stacks comprising all satellites, i.e. not segregated
with respect to the mass of the primary galaxy.
}
\label{fig:Hbeta}
\end{figure}

\section{Stacking procedure}
\label{Sec:Method}

Following \citet{GAMA:CPs}, the SDSS spectra are stacked following two
main parameters, one describing the ``local'' driver of formation, and
a second one related to the presence of the pair. For the former, we
choose the velocity dispersion, and for the latter, we choose the
stellar mass ratio, $\mu$, as defined above. We use the SDSS official estimates of velocity
dispersion, as provided by the DR14 
\citep{SDSSDR14} {\sc SpecObj} catalogue (parameter velDisp). 
Note that, in contrast to
this work, \citet{GAMA:CPs} used the stellar mass of the satellite
galaxy instead of velocity dispersion. The main reason to choose
stellar mass was the inherently larger uncertainty in the estimate of
velocity dispersion because of the lower S/N of the spectra.
However, the velocity dispersion correlates strongly
with the population
properties such as observed colour, age, metallicity or [$\alpha$/Fe]
\citep[see, e.g.,][]{Bernardi:03,Thomas:05,Graves:09,Nic:17,Barone:18},
and provides a better tracer of the underlying stellar populations,
whereas the stellar mass is not so strongly correlated with population
properties.  In the discussion section, we will elaborate on the
differences between these two choices of a local driver.

\begin{figure*}
\begin{center}
\includegraphics[width=85mm]{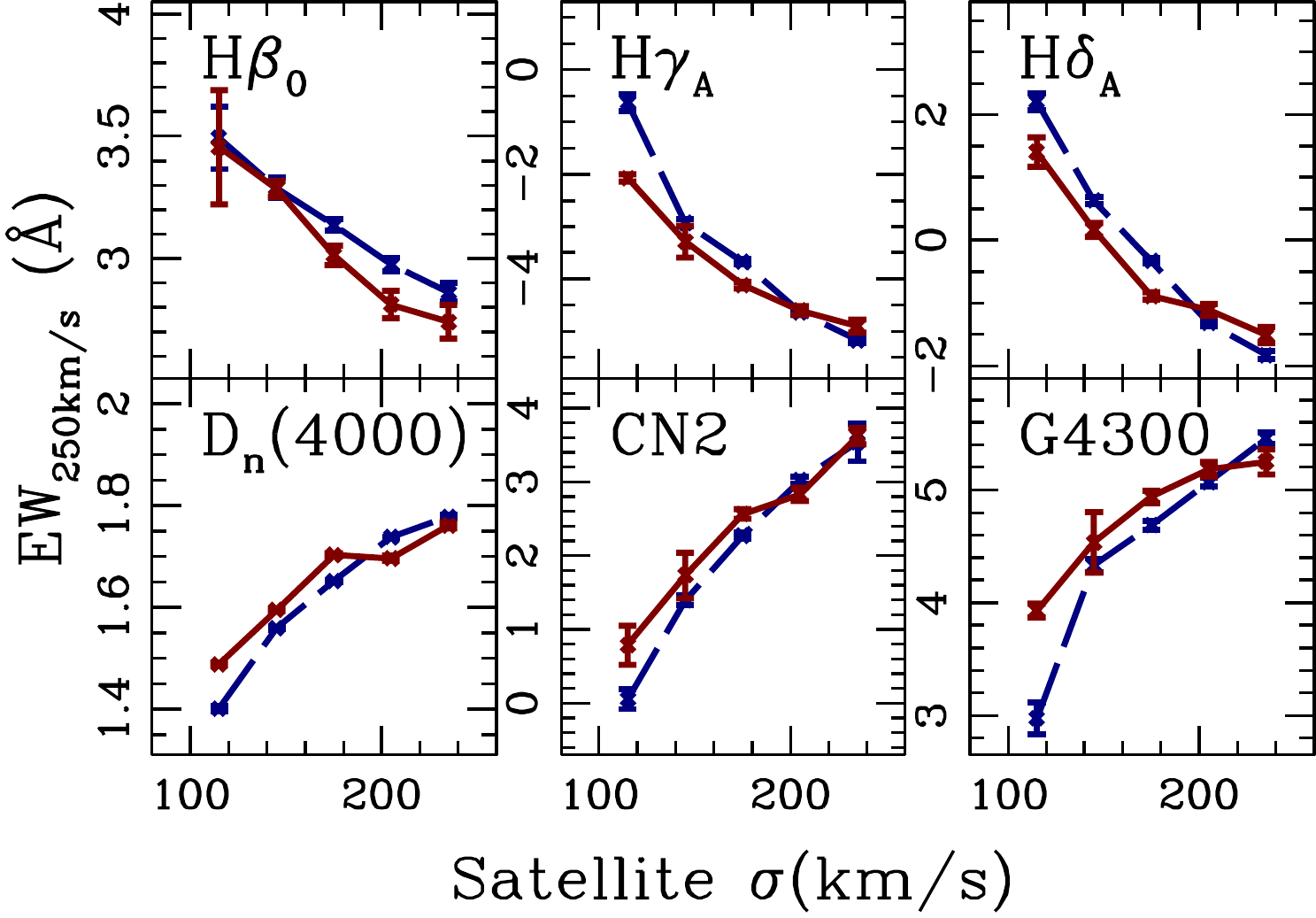}
\includegraphics[width=85mm]{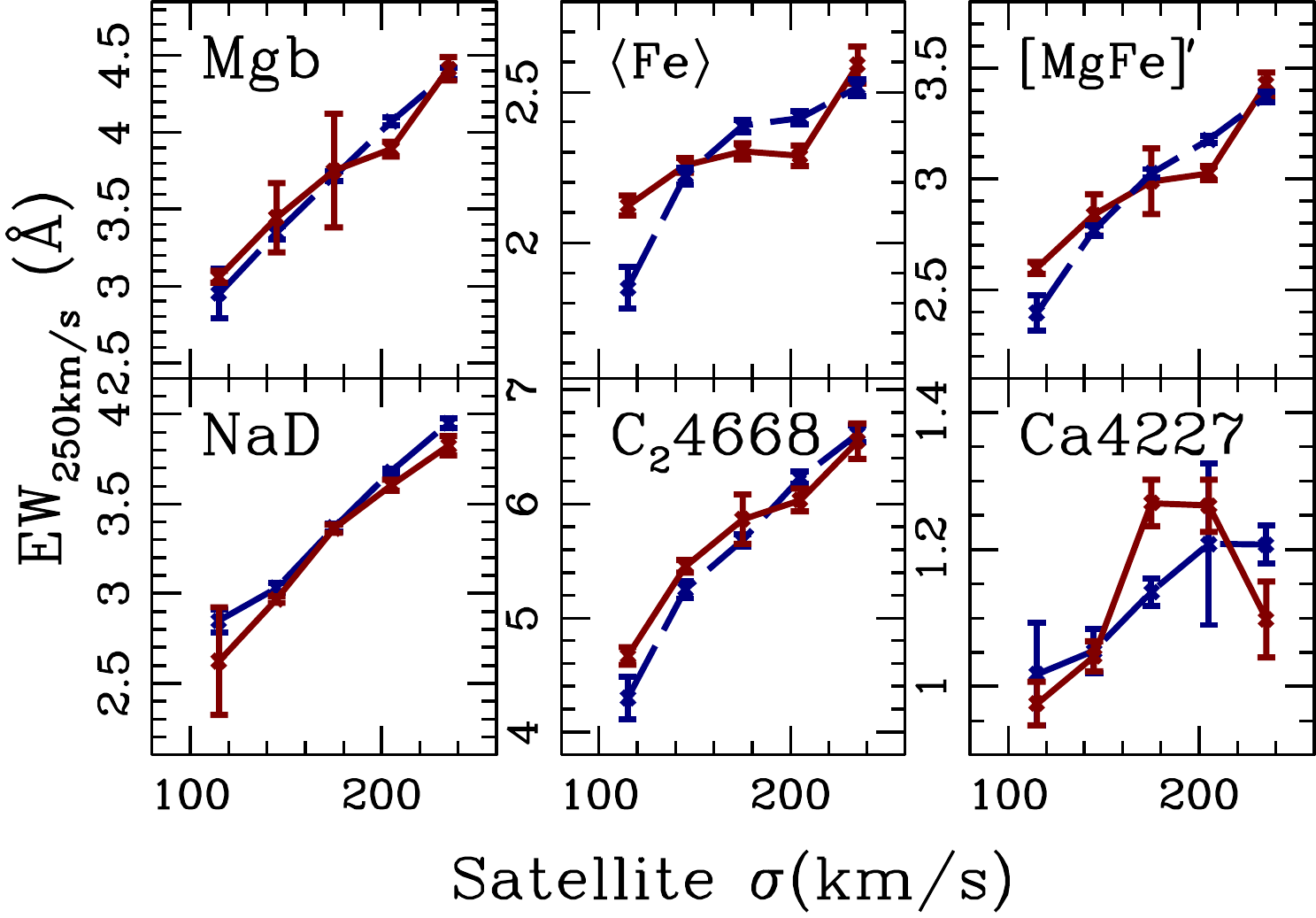}
\end{center}
\caption{Equivalent widths of a number of
age- (LHS) and metallicity-sensitive (RHS) spectral features, measured on the
stacks. The sample is shown with respect to the measured velocity
dispersion of the satellite galaxy (horizontal axis), but the spectra
are convolved to a common dispersion of 250\,km\,s$^{-1}$, to assess
potential differences. The solid red (dashed blue) lines correspond to satellites
around the most (least) massive primaries.
}
\label{fig:EWs}
\end{figure*}

All spectra are corrected for foreground (Milky Way) dust extinction
-- following the standard extinction law of \citet{CCM:89}, with the
colour excess determined from the dust maps of \citet{Schlafly:11} --
and brought into the rest frame. The spectra are normalized in flux in the
wavelength interval 5,000-5,500\AA\ and the resulting flux is drizzled to a
reference grid, performing linear interpolation to split the flux
between adjacent bins. Two sources of uncertainty must be taken into account:
on the one hand, we propagate the statistical uncertainties of
individual flux measurements -- given by the inverse variance in each
entry of the SDSS spectra. Moreover, additional scatter will be
expected from the stacking procedure, due to intrinsic variations in
the properties of the galaxies whose spectra are stacked. To quantify
the latter source of uncertainty, we perform a bootstrap method
whereby 100 realizations of each stack are created by selecting each
time a random set comprising 75\% of the original galaxies within a
given subsample (i.e. for a choice of velocity dispersion and stellar
mass ratio bin). The resulting standard deviation is added in
quadrature to the statistical scatter from individual measurements.
We note that the drop in S/N caused by adding this second source of
noise (from variance in the galaxy sample within a bin) stays within 
10-20\% of the `intrinsic' noise obtained from the spectra. This is
an important issue, confirming that the galaxy-to-galaxy variance
within a velocity dispersion bin does not dominate the error budget.
The total S/N for each stack is 
quoted in Table~\ref{tab:stacks}, and given as an average
in the rest-frame 5,000-5,500\AA\ spectral window.  Moreover, the
difference between the final flux values and those that correspond to
the median of the distribution of 100 realizations stays below 20\% of
the final error. Our errors are thus conservative.

For the comparison among all bins, 
we decided to bring all the stacked spectra to a common velocity
dispersion. Although this technique doubtlessly washes out
information from the spectra, our aim is to robustly constrain
differences in the populations of satellite galaxies, at fixed stellar
mass, with respect to the mass ratio of the pair. Given the range of
velocity dispersion of the sample, we chose
$\sigma_0=250$\,km\,s$^{-1}$, as the common value. To do that, we
convolve the stacks with a Gaussian kernel, inspecting the result
with {\sc pPXF} \citep{pPXF} until the output velocity dispersion matches the
targeted $\sigma_0$.  We note the process produces the same, fiducial,
velocity dispersion with an uncertainty less than $\sim 2$\,km\,s$^{-1}$.

To illustrate the validity of the stacked spectra as a true
representative of a subsample, we show in Fig.~\ref{fig:Dn4K} the
distribution of the D$_n$(4000) index \citep[as defined in ][]{Dn4000}
measured in {\sl individual} spectra, compared with the equivalent
index measured in the stacked spectra (dots) corresponding to each
subsample, shown with respect to velocity dispersion, in increasing
order from the top down. The histograms are smoothed
following a standard Gaussian Kernel Density Estimator \citep[see,
e.g.,][]{Chen:17}.  In each panel, the symbols show, from top to
bottom, the measurement of D$_n$(4000) on the stacks, the median, and
the mean of the individual measurements, respectively. The blue open
and red solid dots represent the $\mu_1$ and $\mu_3$ sets,
respectively. Despite the substantial scatter in the individual
measurements, the trends follow that of the stacked results, thus
justifying the use of stacked spectra to compare the subsamples.  Note
that noisy data will tend to reduce the break strength.

The next step involves removing the contribution from nebular
emission. This is especially important in the Balmer lines, because we will
be targeting H$\beta$, H$\gamma$ and H$\delta$ as key line strength
indicators regarding stellar ages. We ran the spectral fitting code
{\sc STARLIGHT} \citep{Starlight}, which performs an MCMC search for a
best fit, using mixtures of simple stellar populations. Our base grid
comprises a set of 176 SSP spectra from the MILES population synthesis
models \citep{MIUSCAT}, with 44 age steps between 0.1 and 14.1\,Gyr
and four metallicity bins, namely [Z/H]=\{-0.71,-0.40,0.00,+0.22\}.
The spectra are fit in the region 3,750-7,000\,\AA.  The code includes
dust attenuation as an additional parameter, and we made sure the
emission line regions were masked out during the fits. The output
best-fit spectra are compared with the original one, and a Gaussian
fit is performed in eleven spectral regions corresponding to Balmer
H$\alpha$ to H$\delta$, [OIII] at 4959 and 5007\AA\ , [NII] at 6548
and 6583\AA\ , and [SII] at 6716 and 6731\AA .  We use these best-fit
Gaussians to remove the flux from the original spectra, and the line
strengths are measured on the cleaned spectra.  Fig.~\ref{fig:Hbeta}
plots the equivalent width of the emission line correction in the H$\beta$
line as a function of velocity dispersion, with the dashed blue line
(solid red line) representing the $\mu_1$ ($\mu_3$) stacks. The grey
line shows the result for the $\mu_0$ sample (i.e. no segregation with
respect to the mass ratio). Note the monotonically decreasing trend of
the emission component with velocity dispersion, and the significantly
higher level of emission in the $\mu_1$ stacks, corresponding to
satellites around less massive primaries.

Fig.~\ref{fig:EWs} shows the line strengths of the cleaned spectra at
the fiducial 250\,km\,s$^{-1}$ velocity dispersion, in the $\mu_1$
(dashed blue lines) and $\mu_3$ (solid red lines) subsamples, with
respect to the velocity dispersion of each bin.
The indices used in the analysis consist of the standard age-sensitive
indicators: H$\beta_o$ \citep{Hb0}, H$\gamma_A$ and
H$\delta_A$ \citep{Hgd}, D$_n$(4000) \citep{Dn4000}, CN2 and G4300
\citep{Trager:98}. We complement the analysis with a set of metallicity-sensitive
indicators: Mgb, $\langle$Fe$\rangle\equiv$Fe5270+Fe5335, NaD, C$_2$4668, 
and Ca4227 \citep{Trager:98} and [MgFe]$^\prime$ \citep{MgFep}.  The
data show a consistent local trend from young, possibly metal-poor
populations at low velocity dispersion, towards an older, metal-rich
composition in the more massive stacks.  In addition to this locally-driven 
trend, we find a consistent environment-related trend, such that {\sl
at fixed satellite velocity dispersion}, the $\mu_1$ sample -- involving
satellites where the mass ratio $\mu$ is closer to 1:1, therefore
associated to the lowest mass primaries -- feature younger
populations. This trend is very similar to the one found in
\citet{GAMA:CPs}, who used a different set of spectra assembled from GAMA/AAT data. 
Moreover, note this behaviour mirrors that of the emission lines in
Fig.~\ref{fig:Hbeta}, as the stacks with younger populations also
produce higher emission line corrections. This result appears to be
quite robust, especially considering that the age-sensitive indices
D$_n$(4000), CN2 and G4300 are independent of any nebular emission
line correction. Such a trend is quite remarkable, because it shows
that the stellar population ages of satellite and primary galaxies in
pairs are linked. The cause of such a link is discussed in
Section~\ref{Sec:Disc}. In the next section, we will translate these
line strength differences into stellar age trends.

\section{Extracting SSP-equivalent parameters}

When translating the observed line strength differences into
variations of the stellar populations, we decided to keep the
potential trends as clear cut as possible in this paper. We opted to
work with SSP-equivalent variations, namely the observed line
strengths are compared with a large volume of single stellar
populations over a wide range of ages and metallicities.  A more
complex set of models based on extended star formation histories
complicates the analysis beyond the scope of this paper. Our
fundamental aim is to assess whether -- in line with our previous
findings based on AAT spectra in the GAMA survey -- significant
differences are found between satellite galaxies with the same
velocity dispersion (i.e. the ``local driver''), caused by the
presence of a nearby massive primary (i.e. the ``environment
driver''). An SSP-equivalent derivation is not only satisfactory for
our purposes, but the quantification of the sought differences are
better defined than with extended formation models. However, in order
to ascertain that the derived variations are not produced by a
model-related systematic, we will consider two completely independent
sets of population synthesis models. Our data will be fitted with the
standard BC03 models of \citet{BC03} and the more recent MIUSCAT
models of \cite{MIUSCAT}. Not only do these models have different
implementations of the isochrones, but they are also based on
different stellar libraries.

\begin{table}
\caption{SSP-equivalent ages of satellite galaxies (in Gyr), including the
68\% confidence level. The meaning of the $\mu$ and $\sigma$ subsets is
shown in Table~\ref{tab:stacks}}
\label{tab:ages}
\begin{tabular}{c|ccccc}
\hline
ID & $\sigma_1$ & $\sigma_2$ & $\sigma_3$ & $\sigma_4$ & $\sigma_5$\\
\hline
\multicolumn{6}{c}{BC03 models}\\
\hline
$\mu_1$ & $1.71_{-0.48}^{+0.23}$ & $2.25_{-0.38}^{+0.52}$ & $2.70_{-0.43}^{+0.39}$ & $3.37_{-0.44}^{+0.80}$ & $4.42_{-1.05}^{+1.44}$\\
$\mu_2$ & $1.94_{-0.53}^{+0.44}$ & $2.27_{-0.40}^{+0.52}$ & $3.19_{-0.59}^{+0.84}$ & $4.80_{-1.06}^{+1.45}$ & $3.46_{-0.46}^{+0.77}$\\
$\mu_3$ & $1.93_{-0.48}^{+0.45}$ & $2.56_{-0.49}^{+0.50}$ & $2.99_{-0.47}^{+0.61}$ & $3.50_{-0.59}^{+0.83}$ & $3.84_{-0.68}^{+1.13}$\\
\hline
$\mu_0$ & $1.82_{-0.48}^{+0.45}$ & $2.31_{-0.41}^{+0.55}$ & $3.01_{-0.53}^{+0.69}$ & $3.54_{-0.58}^{+0.86}$ & $3.70_{-0.57}^{+0.89}$\\
\hline
\multicolumn{6}{c}{MIUSCAT models}\\
\hline
$\mu_1$ & $1.67_{-0.18}^{+0.19}$ & $2.35_{-0.32}^{+0.32}$ & $2.81_{-0.31}^{+0.35}$ & $3.60_{-0.41}^{+0.91}$ & $4.68_{-0.81}^{+0.97}$\\
$\mu_2$ & $2.22_{-0.42}^{+0.29}$ & $2.49_{-0.39}^{+0.45}$ & $3.24_{-0.41}^{+0.68}$ & $4.82_{-1.09}^{+1.14}$ & $3.97_{-0.52}^{+1.13}$\\
$\mu_3$ & $2.19_{-0.33}^{+0.20}$ & $2.52_{-0.40}^{+0.53}$ & $3.41_{-0.40}^{+0.78}$ & $4.02_{-0.71}^{+1.27}$ & $3.89_{-0.44}^{+1.14}$\\
\hline
$\mu_0$ & $2.05_{-0.35}^{+0.25}$ & $2.41_{-0.40}^{+0.52}$ & $3.22_{-0.40}^{+0.72}$ & $4.13_{-0.71}^{+1.17}$ & $4.19_{-0.57}^{+1.09}$\\
\hline
\end{tabular}
\medskip
\end{table}

The ages are derived by comparing the observed line strengths with a
grid of SSP models with ages ranging from 0.5 to 14\,Gyr in 256 (logarithmic)
steps, and from [Z/H]=$-$0.5 to +0.3 in 64 steps (the MIUSCAT grid only 
extends out to +0.22 in [Z/H] as this is the highest available value
of metallicity). Both sets adopt a fiducial, Milky Way-like initial
mass function (\citealt{Chabrier:03} for the BC03 models, and \citealt{KuIMF}
for MIUSCAT).  The line strengths of the
model grid are compared with the observations with a standard $\chi^2$
statistic. However, at the high S/N of the stacks (see Table~\ref{tab:stacks}), a naive comparison
of the line strengths will not be capable of fitting all features
consistently.  However, we aim at looking for {\sl relative}
differences between the stacked spectra. To achieve this goal, we
define a fiducial stack -- corresponding to ($\sigma_1$,$\mu_0$) --
introducing offsets to the line strengths, so that the fitting
procedure gives an acceptable reduced $\chi^2$ to the fiducial
stack. The offsets thus produced, are applied to the rest of the
stacks for the analysis. Moreover, these offsets are also added in
quadrature to the individual line strengths, so that less weight is
given to those indices that require larger modifications.  Even though
the absolute values of age and metallicity are not to be trusted with
this methodology, the relative variations should be robust (when
interpreted as SSP-equivalent differences).

\begin{figure*}
\begin{center}
\includegraphics[width=75mm]{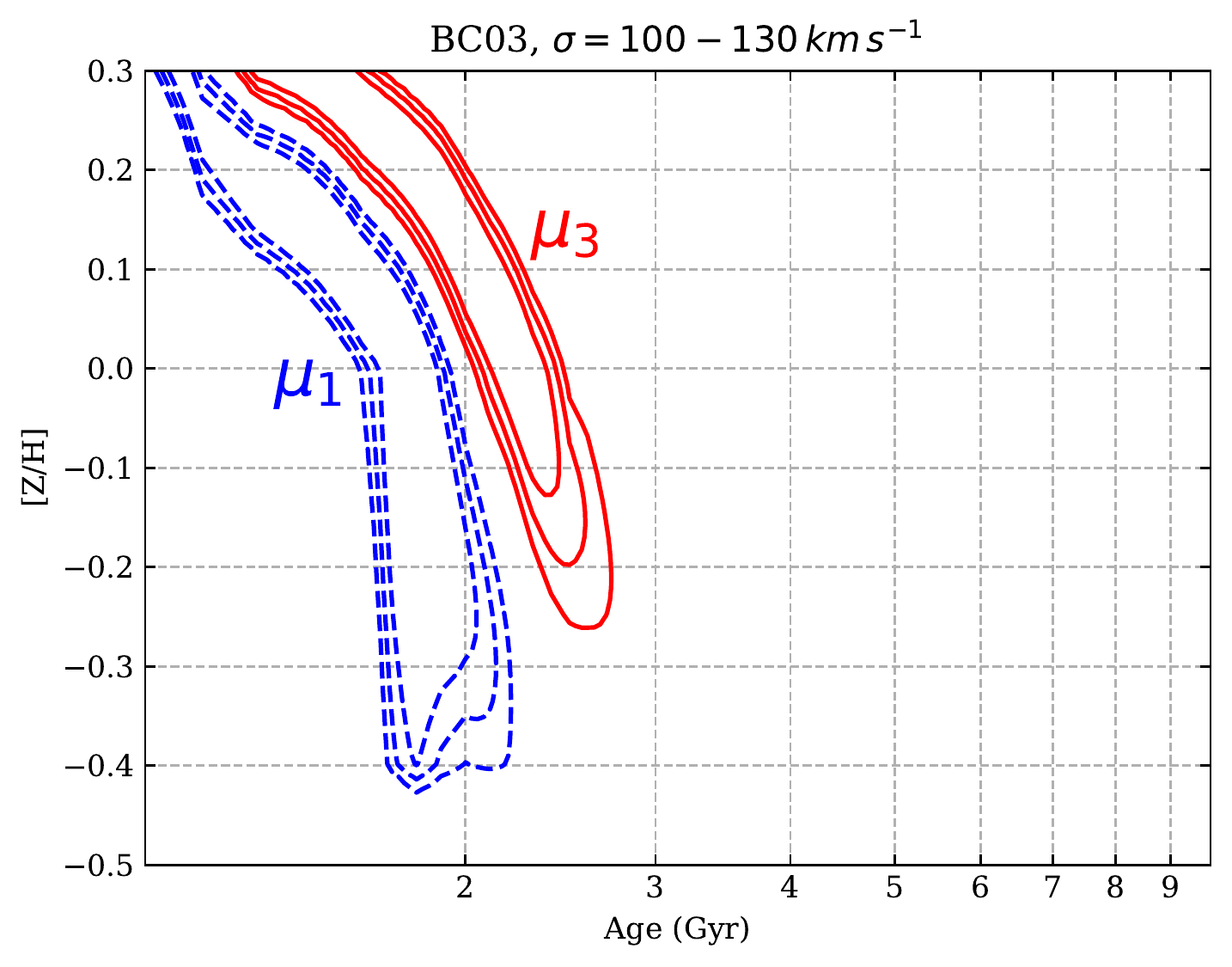}
\includegraphics[width=75mm]{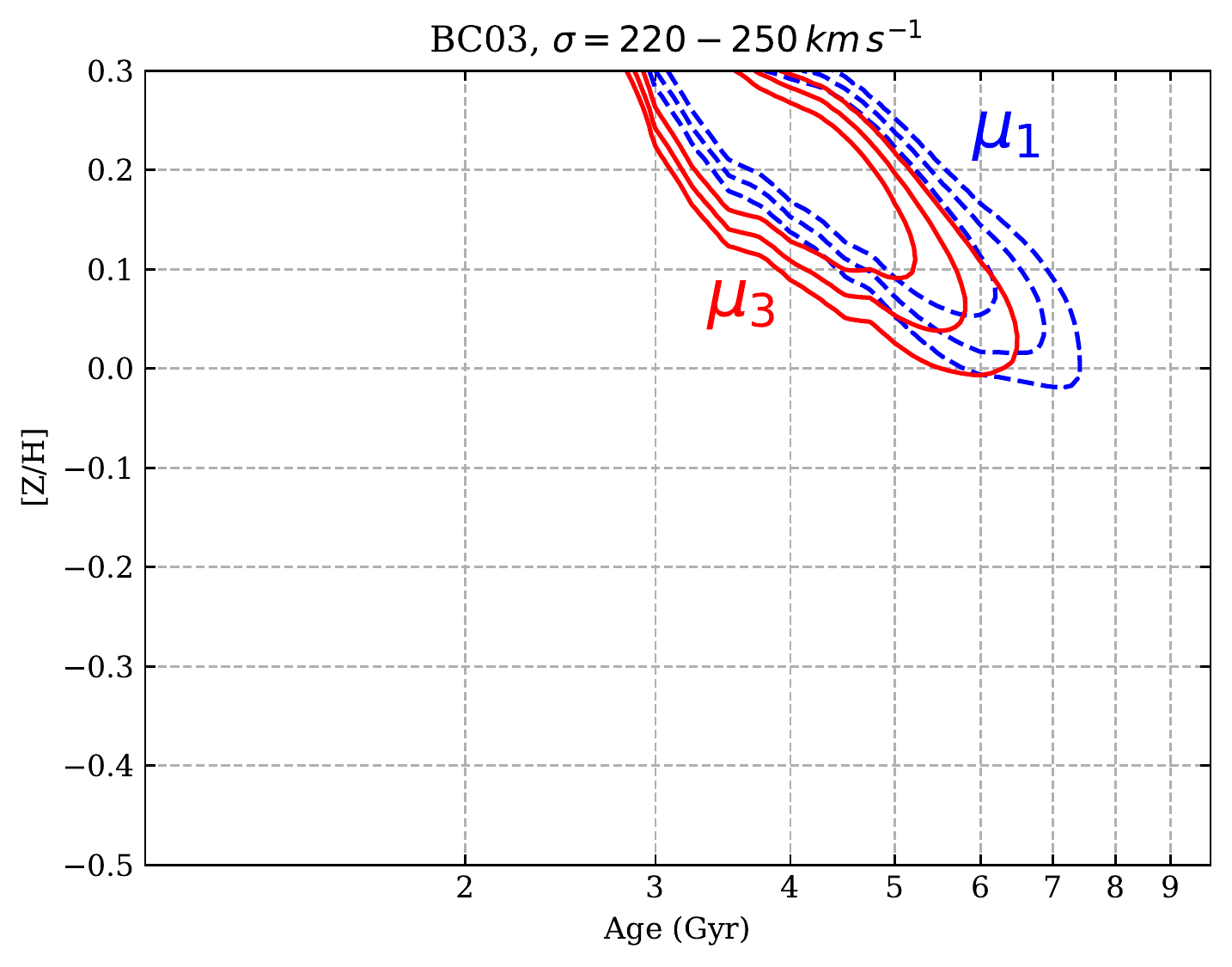}\\
\includegraphics[width=75mm]{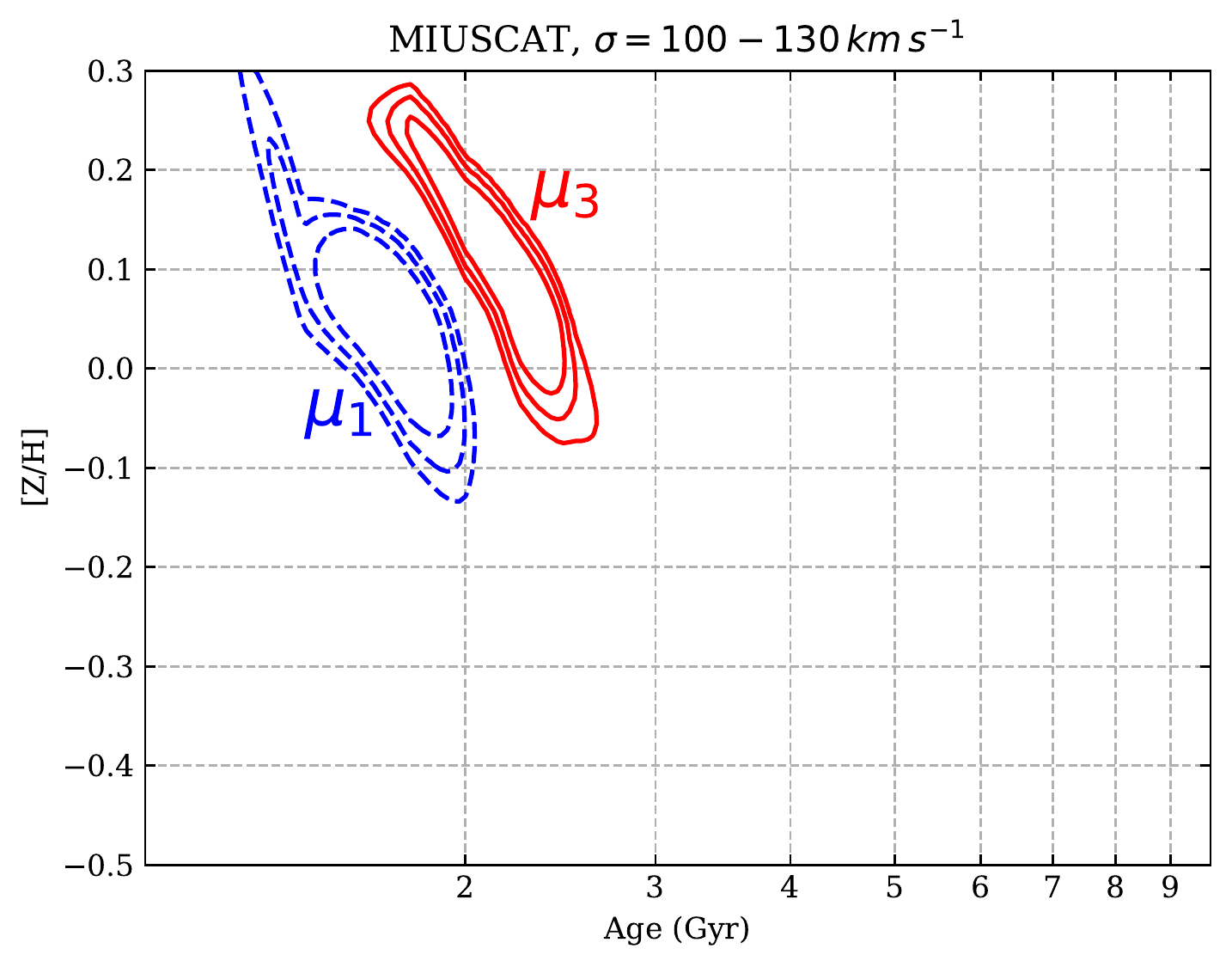}
\includegraphics[width=75mm]{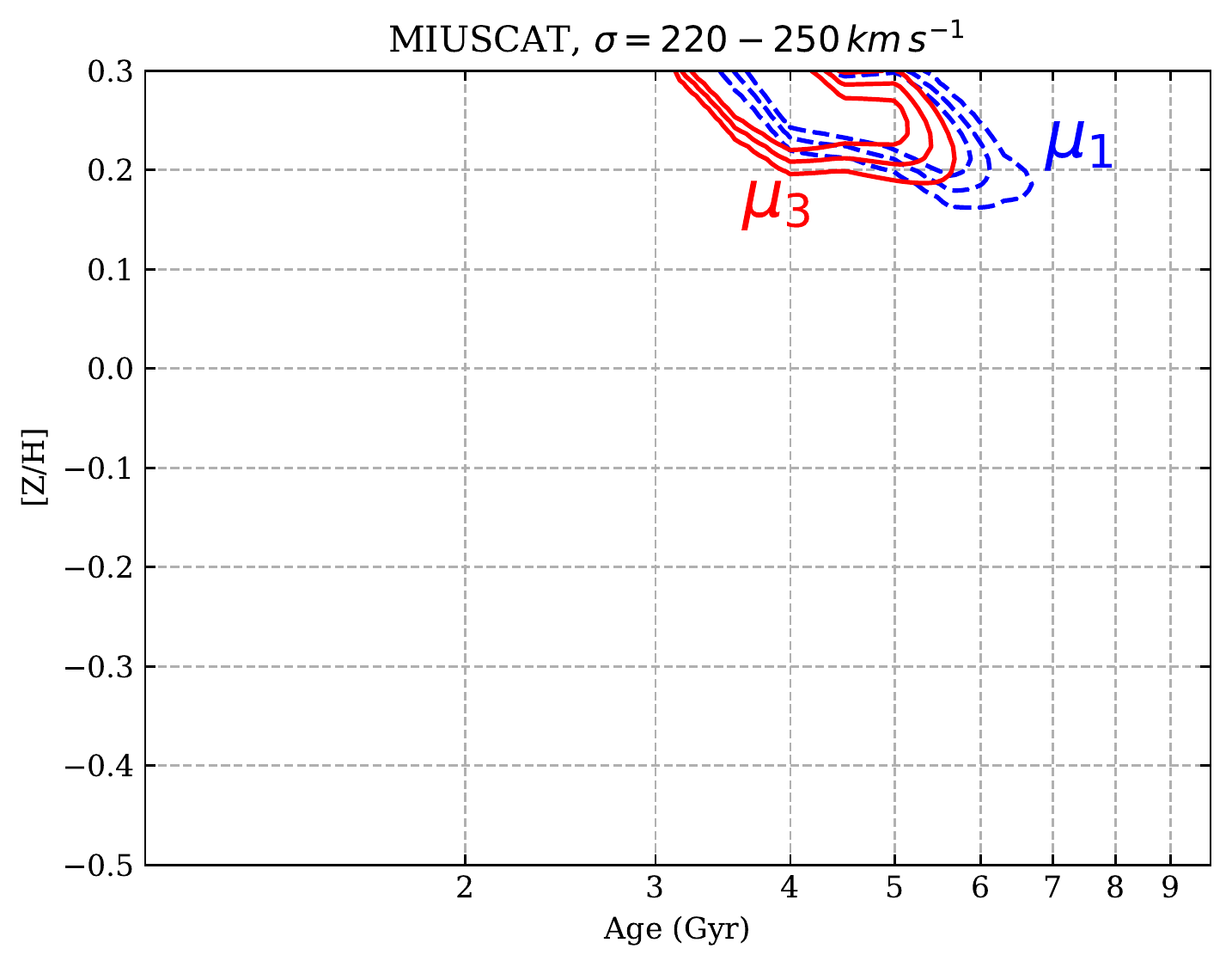}
\end{center}
\caption{Probability maps of the SSP-equivalent age and metallicity.
The contour levels are shown at the 1, 2, and 3\,$\sigma$ confidence
levels (from the inside out) in the $\mu_1$ (blue dashed lines) and
the $\mu_3$ (solid red lines) subsamples. Two different sets of
population synthesis models are used: \citet[][top panels]{BC03}
and MIUSCAT \citep{MIUSCAT} models (bottom panels). The two extreme
choices of velocity dispersion are shown, as labelled.
}
\label{fig:c2maps}
\end{figure*}

\begin{figure}
\begin{center}
\includegraphics[width=85mm]{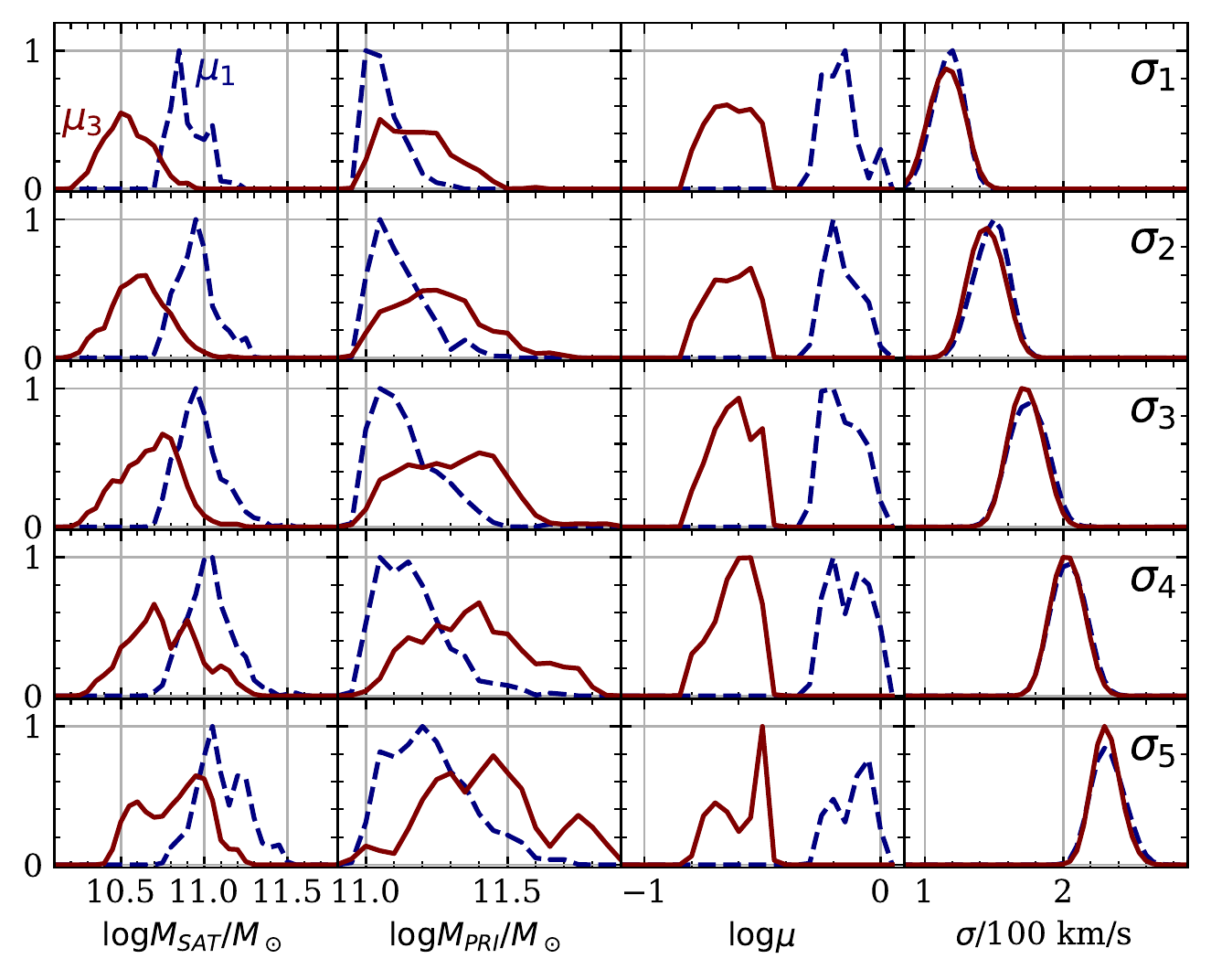}
\end{center}
\caption{Distribution of subsample properties in the
$\mu_1$ (blue dashed histograms) and $\mu_3$ subsets (solid
red histograms). The sample is binned with respect to
velocity dispersion, as labelled, in increasing order
of $\sigma$ from the top down. See Tables~\ref{tab:stacks}
and \ref{tab:stacks_info} for more details.}
\label{fig:Hists}
\end{figure}

Table~\ref{tab:ages} shows the SSP-equivalent ages for each of the
stacks, with the confidence levels quoted at 1\,$\sigma$. These
results are marginalized with respect to metallicity (using a flat
prior over the range of metallicity considered). A clearer
assessment of differences with respect to the stellar mass ratio can
be seen in Fig.~\ref{fig:c2maps}, where we show the full 2D
probability distribution corresponding to the analysis (from the
inside out, the contour
plots represent the 1, 2 and 3\,$\sigma$ confidence levels). While the
ages quoted in Table~\ref{tab:ages} do not show so clearly the difference between
stacks $\mu_1$ and $\mu_3$ at fixed velocity dispersion, the figure
strongly suggests that $\mu_3$ stacks (i.e. satellites around the most
massive hosts) are subtly, but consistently older than their
counterparts, {\sl at fixed velocity dispersion}, around lower mass
primaries. This effect is most noticeable at low velocity dispersion,
with differences in stellar age around $\sim$0.5\,Gyr, a result consistent
with the study based on the GAMA/AAT sample of \citet{GAMA:CPs}. However, we
need to describe in some detail the differences between the sample selection
performed here -- where velocity dispersion is the main stacking parameter --
and the selection adopted in that paper -- where stellar mass is the
main stacking parameter. Fig.~\ref{fig:Hists} shows the distribution of
galaxies in the $\mu_1$ (dashed blue) and $\mu_3$ (solid red)
subsamples used in this paper, with respect to several observables, as
labelled. The mean and standard deviation of the distributions are
quantified in Table~\ref{tab:stacks_info}. Note the expected trivial
behaviour of the data with respect to velocity dispersion (rightmost
panels). The panels concerning the distribution in the mass ratio
($\log\mu$) also show a clear separation between the $\mu_1$ and
$\mu_3$ subsets. Note that the histograms with respect to either
satellite or primary stellar mass reveal that the difference between
these two subsets lies in a complex mixture of satellite and primary
masses, such that the $\mu_3$ (older) subsample consistently
represents a distribution where the primary mass is higher than in
$\mu_1$, but also the satellite mass is {\sl lower} than that of the
$\mu_1$ (i.e. younger) subsample (all at fixed velocity dispersion).
We emphasize that velocity dispersion correlates more strongly with
the population parameters than stellar mass, and is less prone to
systematic uncertainties. Therefore, the results in this paper are
more robust than those based on stellar mass, presented
in \citet{GAMA:CPs}.  The fact that satellites with lower mass are
{\sl older} than the more massive counterparts, when orbiting more
massive primaries illustrates the significance of this
environment-related trend. Also note that the trend cannot be ascribed to
a bias caused by the flux limit imposed on the parent SDSS spectroscopic
sample. Such a bias would mainly affect the $\mu_3$ subset (i.e.
lower mass satellites), so that fainter galaxies with the same
velocity dispersion -- therefore older -- would be missed from
the stacks. In this case, the analysis would result in even
older stellar ages of satellites in the $\mu_3$ stack, implying that
the age difference found here is resilient against incompleteness from the 
flux-limited sample selection.

\begin{table}
\caption{Mean and standard deviation of the distribution
of a few properties of the individual galaxies in each
stack (see Table~\ref{tab:stacks} for the meaning of the
$\mu$ and $\sigma$ subindices). The full distributions
are presented in Fig.~\ref{fig:Hists}.}
\label{tab:stacks_info}
\centering
\begin{tabular}{c|ccc}
\hline
 & $\mu_1$ & $\mu_2$ & $\mu_3$\\
\hline
& \multicolumn{3}{c}{$\log(M_{\rm SAT}/M_\odot)$}\\
\hline
$\sigma_1$ & $10.87\pm 0.10$ & $10.73\pm 0.12$ & $10.53\pm 0.14$\\
$\sigma_2$ & $10.94\pm 0.12$ & $10.77\pm 0.14$ & $10.61\pm 0.16$\\
$\sigma_3$ & $10.96\pm 0.13$ & $10.81\pm 0.12$ & $10.69\pm 0.17$\\
$\sigma_4$ & $11.03\pm 0.14$ & $10.87\pm 0.14$ & $10.73\pm 0.21$\\
$\sigma_5$ & $11.08\pm 0.15$ & $10.91\pm 0.16$ & $10.84\pm 0.19$\\
\hline
& \multicolumn{3}{c}{$\log(M_{\rm PRI}/M_\odot)$}\\
\hline
$\sigma_1$ & $11.05\pm 0.07$  & $11.12\pm 0.11$  &  $11.17\pm 0.12$\\
$\sigma_2$ & $11.10\pm 0.10$  & $11.15\pm 0.13$  &  $11.24\pm 0.15$\\
$\sigma_3$ & $11.12\pm 0.11$  & $11.19\pm 0.13$  &  $11.31\pm 0.17$\\
$\sigma_4$ & $11.15\pm 0.12$  & $11.25\pm 0.14$  &  $11.38\pm 0.18$\\
$\sigma_5$ & $11.21\pm 0.14$  & $11.30\pm 0.15$  &  $11.43\pm 0.20$\\
\hline
& \multicolumn{3}{c}{$\log(M_{\rm SAT}/M_{\rm PRI})$}\\
\hline
$\sigma_1$ & $-0.17\pm 0.07$ & $-0.41\pm 0.06$ & $-0.64\pm 0.09$\\
$\sigma_2$ & $-0.18\pm 0.07$ & $-0.39\pm 0.06$ & $-0.63\pm 0.09$\\
$\sigma_3$ & $-0.18\pm 0.08$ & $-0.38\pm 0.06$ & $-0.62\pm 0.09$\\
$\sigma_4$ & $-0.13\pm 0.08$ & $-0.40\pm 0.06$ & $-0.60\pm 0.09$\\
$\sigma_5$ & $-0.10\pm 0.08$ & $-0.36\pm 0.06$ & $-0.60\pm 0.10$\\
\hline
& \multicolumn{3}{c}{$\sigma/100$\,km\,s$^{-1}$}\\
\hline
$\sigma_1$ & $1.19\pm 0.08$ & $1.17\pm 0.08$ & $1.16\pm 0.09$\\
$\sigma_2$ & $1.50\pm 0.09$ & $1.46\pm 0.08$ & $1.44\pm 0.09$\\
$\sigma_3$ & $1.74\pm 0.09$ & $1.74\pm 0.08$ & $1.72\pm 0.09$\\
$\sigma_4$ & $2.04\pm 0.09$ & $2.04\pm 0.08$ & $2.03\pm 0.08$\\
$\sigma_5$ & $2.31\pm 0.08$ & $2.29\pm 0.08$ & $2.31\pm 0.07$\\
\hline
\hline
\end{tabular}
\end{table}

\begin{figure}
\begin{center}
\includegraphics[width=8.3cm]{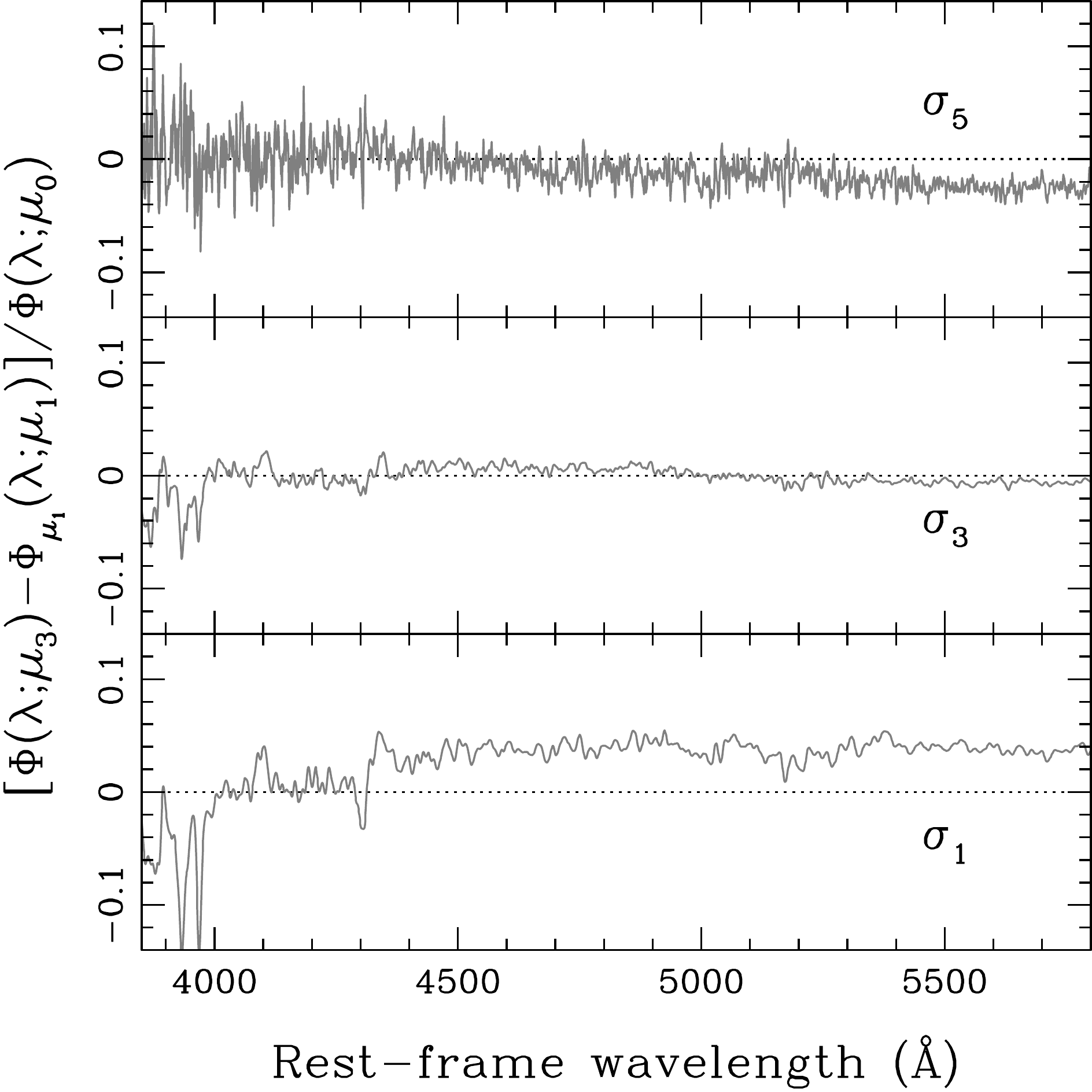}
\end{center}
\caption{Spectral difference between the stacks for $\mu_3$ and $\mu_1$
at three values of the velocity dispersion, as labelled (see
Table~\ref{tab:stacks} for the interpretation of the $\mu$ and $\sigma$
subsamples). The difference is quoted as a fraction of the flux in the
general $\mu_0$ set.
Note the significant excess of red light in the $\mu_3$
stack at low velocity dispersion (bottom panel), a result that is
consistent with the older ages obtained in the targeted line strength
analysis, with respect to the $\mu_1$ subsample.  }
\label{fig:seds}
\end{figure}

Further support for this trend can be found in Fig.~\ref{fig:seds},
where we show the difference between the $\mu_3$ and $\mu_1$ stacks at
three choices of velocity dispersion, spanning the full range
explored.  The difference is quoted as a fraction of the flux in the
full sample at a given velocity dispersion, denoted $\mu_0$ (see
Table~\ref{tab:stacks}).  Note the significant excess of red light in
the spectra of the low-mass satellites ($\sigma_1$) associated to the
most massive primaries ($\mu_3$). This result is fully consistent with
the previous analysis of the line strengths, and confirms the
environment-driven stellar age difference.  In more detail, note the
slight bumps in the $\sigma_1$ set (bottom panel) at the position of
the Balmer lines H$\delta$ (4,100\AA) and H$\gamma$ (4,340\AA), as
expected if $\mu_3$ satellites were older than those in the $\mu_1$
subset. Moreover, the dips at the position of the Ca\,{\sc II} H and K
lines ($\sim$3,950\AA), G-band (4,300\AA) and Mg complex at 5,200\AA\
are suggestive of a slightly lower metallicity in $\mu_3$, but the
population analysis presented above (see Fig.~\ref{fig:c2maps})
confirms that the inherent age-metallicity degeneracy weakens all
possible constraints on metallicity. Although much less significant,
it did not escape our attention the inversion of the age trend
at the highest values of velocity dispersion (rightmost panels of Fig.~\ref{fig:c2maps}
and $\sigma_5$ values in Table~\ref{tab:ages}). The age difference is
compatible, within error bars, with no variation, but it is worth noting
that the $\mu_3$ satellites become  {\sl younger} than the
$\mu_1$ set at the highest velocity dispersion.

\section{Discussion}
\label{Sec:Disc}

The main trend of satellite galaxies being older at fixed velocity
dispersion if the mass ratio $\mu$ is smaller, may be, in principle 
surprising and counterintuitive,
as one generally expects the stellar ages of galaxies to decrease with
decreasing mass \citep[e.g.][]{Gallazzi:05,vdSande:18}.
However, our result can be understood in the context of
assembly bias in hierarchical cosmologies
\citep[e.g.][]{ST:04,AvilaR:05,GW:07}, as lower
$\mu$=M$_{\rm SAT}$/M$_{\rm PRI}$ ratios are typically associated to
older groups, in which all the mergers with the massive primary galaxy
have already taken place. This naturally means that groups with low
$\mu$ collapsed earlier than those groups of the same mass with higher
$\mu$ ratios. If that is the case, then it follows that galaxies in
groups that collapsed earlier are older than those in dynamically
younger groups of the same mass. Extreme examples of the correlation
between group age and the central-to-satellite mass ratio are fossil
groups and clusters
\citep[e.g.][]{Jones:03,Zara:16}, in which the most massive
satellite of the system is at least two magnitudes fainter than the
central, brightest galaxy. \citet{DOnghia:05} and \citet{Dariush:10},
among others, showed that in a hierarchical universe,
these fossil groups are expected to have been formed at much higher
redshift than other groups of the same mass.

Recently, \citet{Zehavi:18} used several cosmological semi-analytic
models of galaxy formation to study the effect of halo assembly bias
on the formation of galaxies and found that at fixed halo mass, older
halos (which collapsed earlier) tend to have a smaller number of
satellites and with satellite-to-central galaxy mass ratios that are
much smaller than younger halos, in line with our findings.
\citet{Artale:18}, using the cosmological hydrodynamical EAGLE and Illustris
simulations (\citealt{Schaye:15} and \citealt{Vogelsberger:14}, 
respectively) showed that the central galaxies of older halos are more
massive than those in younger halos at fixed halo mass by 0.1--0.3\,dex.
Similar differences in stellar mass were reported by 
\citet{Zehavi:18}. This agrees with what we find in our sample, as the
central galaxies of the $\mu_3$ selection are more massive than those
in the $\mu_1$ set by $\sim$0.2\,dex. We should also note that the
recent analysis of \citet{Davies:19} showed that the passive fraction
of satellites increases steeply with decreasing M$_{\rm SAT}$/M$_{\rm
PRI}$ in GAMA, supporting the idea that group age affects the ages and
passive fractions of satellite galaxies. All this evidence points
towards assembly bias being the plausible origin for the trends we
observe. However, to confirm this, simulations would need to mimic our
selection, which is based on velocity dispersion of the satellite
galaxies. We should emphasize that we refer here 
to one-halo assembly bias, i.e. concerning galaxies within the
same dark matter halo. Alternatively, two-halo assembly bias
\citep[see, e.g.,][]{Kauff:13,Sin:17,Tinker:18} affects the properties
over larger scales (beyond 1\,Mpc), and is not related to the claimed effect 
\citep[see the review of][especially their section 6.2, for more
details on this distinction]{WT:18}.

\section{Summary}
\label{Sec:Summ}

The properties of close pairs involving a massive galaxy can be
used to understand the growth mechanisms via mergers, as well as
the role of environment in galaxy formation. This paper extends
the work of \citet{GAMA:CPs}, based on GAMA/AAT spectra, to
a different sample of high S/N spectra from the Sloan Digital Sky
Survey. The methodology is very similar to that paper, selecting
nearby companions of massive galaxies, in  dynamical interaction
leading to a potential merger. Here we use the stellar velocity dispersion
of the satellite as a ``local'' proxy,
and the satellite-to-primary mass ratio, $\mu$, to characterize
the close pair. Very high S/N spectra are created by stacking the
data in a set of five bins in velocity dispersion and three bins
in mass ratio, with a careful selection of redshift, to avoid
systematic trends (Fig.~\ref{fig:sample}). The working sample comprises
about two thousand high quality SDSS spectra -- in the redshift
window 0.07$<$z$<$0.14 -- covering a velocity dispersion
between 100 and 250\,km\,s$^{-1}$ and a satellite-to-primary stellar
mass ratio between 1:6 and 1:1 (Table~\ref{tab:stacks}).

A battery of spectral line strengths is studied to assess the
difference in stellar age between satellites involving different
values of the mass ratio, at fixed velocity dispersion (Fig.~\ref{fig:EWs}). Were
environment-related processes irrelevant, we would have found no
difference in the underlying populations of the stacks with respect to
the mass ratio, $\mu$.  In agreement with our GAMA-based analysis, we
find a consistent trend, such that satellites around the most massive
galaxies are systematically older (based on SSP-equivalent ages, see
Fig.~\ref{fig:c2maps} and Table~\ref{tab:ages}), with the age
difference increasing towards decreasing velocity dispersion, up to
0.5\,Gyr at the lowest velocity dispersions probed
($\sigma\sim$100\,km\,s$^{-1}$).  This result provides yet another
supporting argument of galactic conformity \citep{Weinmann:06} and the
idea of a galaxy assembly bias \citep{Hearin:15}, such that satellites
with low values of the mass ratio $\mu$, are expected to lie in halos
that form earlier, akin to a fossil group.

\section*{Acknowledgements}
IF acknowledges support from the AAO through their distinguished
visitor programme, as well as funding from the Royal Society.  Funding
for SDSS-III has been provided by the Alfred P. Sloan Foundation, the
Participating Institutions, the National Science Foundation, and the
U.S. Department of Energy Office of Science. The SDSS-III web site is
http://www.sdss3.org/.


\label{lastpage}

\end{document}